\documentclass[10pt, twocolumn]{article}
\usepackage[utf8]{inputenc}
\usepackage{geometry}
\usepackage{mwe}
\usepackage{afterpage}
\usepackage{amsmath}
\usepackage{amsfonts}
\usepackage{hyperref}
\usepackage{subcaption}
\usepackage[bottom]{footmisc}
\usepackage{tikz}
\usepackage{xfrac}
\usepackage{siunitx}
\usepackage{multirow}
\usepackage{multicol}
\usepackage{float}
\usepackage{pgfplots}
\usepackage{tikz-3dplot}
\usetikzlibrary{calc}
\usetikzlibrary{math}
\pgfplotsset{width=7cm,compat=1.8}

\usepackage{color}
\usepackage{soul}
\usepackage{xcolor}

\makeatletter
\renewcommand{\fnum@figure}{Fig. \thefigure}
\makeatother

\usepackage{sectsty} 
\usepackage{titlesec}
\sectionfont{\fontsize{10}{12}\selectfont}
\subsectionfont{\fontsize{10}{12}\selectfont}
\titlespacing\section{0pt}{10pt plus 0pt minus 0pt}{10pt plus 0pt minus 0pt}
\titlespacing\subsection{0pt}{0pt plus 0pt minus 0pt}{0pt plus 0pt minus 0pt}
\usepackage{indentfirst} 
\titlelabel{\thetitle.\quad}

\usepackage{changepage}

\usepackage{etoolbox} 
\makeatletter
\patchcmd{\@maketitle}{\LARGE}{\fontsize{18}{24}\selectfont}{}{}
\makeatother

\usepackage[affil-it]{authblk} 

\renewcommand\Affilfont{\bfseries\fontsize{12}{14.4}\selectfont}
\makeatletter
\renewcommand\AB@authnote[1]{\textsuperscript{\normalfont\bfseries#1}}
\makeatother
\makeatletter
\renewcommand\AB@affilsepx{, \protect\Affilfont}
\makeatother

\usepackage{fancyhdr}
\usepackage[style=ieee, 
citestyle=numeric-comp,
sorting=none]{biblatex}
\usepackage{enumitem}
\setlist[itemize]{nosep}
\setlist[enumerate]{nosep}

\title{\textbf{Stochastic MPC with Dual Control for Autonomous Driving with Multi-Modal Interaction-Aware Predictions}}
\author[1]{Siddharth H. Nair}
\author[1]{Vijay Govindarajan}
\author[2]{Theresa Lin}
\author[3]{Yan Wang}
\author[3]{Eric H. Tseng}
\author[1]{Francesco Borrelli}
\affil[1]{Department of Mechanical Engineering, University of California, Berkeley}
\affil[2]{Ford AV LLC}
\affil[3]{Ford Research and Advanced Engineering}

\date{\vspace{-8.5ex}}

\begin{document}

\twocolumn[
  \begin{@twocolumnfalse}
    \maketitle
    
    \begin{center}
    Email: \href{mailto:siddharth_nair@berkeley.edu}{\color{blue} \underline{\smash{siddharth\_nair@berkeley.edu}}}
    \end{center}
    
    % the abstract
    \begin{adjustwidth}{15mm}{15mm}
    We propose a Stochastic MPC (SMPC) approach for autonomous driving which incorporates multi-modal, interaction-aware predictions of surrounding vehicles. For each mode, vehicle motion predictions are obtained by a control model described using a basis of fixed features  with unknown weights. The proposed SMPC formulation finds optimal controls which serves two purposes: 1) reducing conservatism of the SMPC by optimizing over   parameterized control laws and 2) prediction and estimation of feature weights used in interaction-aware modeling using Kalman filtering. The proposed approach is demonstrated on a longitudinal control example, with uncertainties in predictions of the autonomous and surrounding vehicles.
    \end{adjustwidth}
    
    % the paper topics
    \begin{center}
    Autonomous Driving Systems, Advanced Driver Assistance Systems, Identification and Estimation
    \end{center}
    
  \end{@twocolumnfalse}
]

\thispagestyle{fancy} % add header
\renewcommand{\headrulewidth}{0pt} % remove header line

\section{INTRODUCTION}
Autonomous driving technologies have seen a surge in popularity over the last decade, with the potential to improve flow of traffic, safety and fuel efficiency \cite{nhtsa}. While existing technology is gradually being introduced into traffic, the absence of V2V communication makes safe motion planning for autonomous agents a challenge. The difficulty arises because of the variability of possible behaviors of the surrounding agents. To address this, works like \cite{ivanovic2020mats, covernet_2020} build prediction models for the surrounding agents with multiple modes (discrete decisions for distinct maneuvers), which may also incorporate interactions with the autonomous agent.   

The focus of this work is to use these multi-modal, interaction-aware predictions of the surrounding agents (called Target Vehicles (TVs)) in a Model Predictive Control (MPC) \cite{borrelli2017predictive} framework for the autonomous agent (called Ego Vehicle (EV)). We use Stochastic MPC (SMPC) for addressing uncertainties in EV and TVs predictions by using probabilistic collision avoidance, and state and actuation constraints. The TVs’ predictions for each mode are obtained using a driver/control model with a  basis of known features\footnote[1]{
E.g., features for a longitudinal control driver model may consist of safety distance from its lead car and/or the distance to a stop sign.} multiplied by  unknown, time-varying weights (e.g., such a model can be obtained by regression \cite{hastie2009elements}).

\section{RELATED WORK}
There is a large body of work focusing on SMPC for autonomous driving applications, e.g., autonomous lane change \cite{carvalho2014stochastic}, cruise control\cite{moser2017flexible}, and platooning\cite{causevic2020information}. 
A typical SMPC algorithm involves solving a chance-constrained finite horizon optimal control problem in a receding horizon fashion \cite{mesbah2016stochastic}. The prevailing approaches (\cite{brito2021learning}) for autonomous driving solve the SMPC optimization problem to find a single sequence of control values to satisfy the collision avoidance constraints for TVs’ trajectory predictions of all the modes. This can be conservative and feasible solutions for such SMPC may not exist, which is undesirable in practice if a backup planner is unavailable. 

We build on \cite{nair2021stochastic} to propose an SMPC scheme which optimizes over a sequence of control laws. This allows for different sequences of control values corresponding to different realizations of the EV’s and TVs’ trajectories, thus enhancing feasibility of the SMPC optimization problem. The SMPC also uses ideas from dual control \cite{mesbah2018stochastic, arcari2020dual} for prediction and estimation of the unknown, feature weights (specific to TV drivers) corresponding to each mode using Kalman filtering. We also maintain a probability distribution over modes using Bayesian updates, which allows prioritising performance for the more probable modes. 

The paper is organized as follows. Section \ref{sec:pf} describes the control problem and modelling assumptions. In section \ref{sec:SMPC}, we detail our SMPC design for addressing the control problem and demonstrate our approach in simulation, along with ablation studies in section \ref{sec:sim}.
\section{PROBLEM FORMULATION}\label{sec:pf}
In this section, we formalize our modeling assumptions and present our SMPC design.
\subsection{EV Prediction Model}
The EV is modeled using a linear time-varying (LTV) system with state $x_t$, input $u_t$ and noise $w_t$ at time $t$, given by eq. (1). The noise is normally distributed with mean $0$, variance $\Sigma_w $ and is i.i.d. for all $t\geq0$.
{\small{
\begin{subequations}
% \vskip -0.25 true in
\begin{align}
    x_{t+1}&=A_tx_t+Bu_t+w_t\\
    w_t&\sim\mathcal{N}(0,\Sigma_w)
\end{align}
% \vskip -0.1 true in
\end{subequations}}}
The state space and system matrices depend on the relevant application and chosen coordinate system.
% \footnote[2]{For longitudinal control, a point-mass model with state $x_t=\left[s_t\ v_t\right]^\top$ consisting of longitudinal position  and speed, and acceleration as input $u_t=a_t$ can be used.}.
\subsection{TV Prediction Model}
The TVs are jointly modeled in eq. (2) with all TVs’ states given by $o_t$, i.i.d. noise $v_t$ and interaction-aware driver/control models given by a feature matrix {\small{$\Phi^\sigma\left(x_t,o_t\right)$}} for a particular mode {\small{$\sigma\in\{1,\ldots,\ M\}$}},  multiplied by weights $\gamma_t$
{\small{
\begin{subequations}\label{eq:tv_dyn}
% \vskip -0.25 true in
\begin{align}
    o_{t+1}&=\bar{A}_to_t+\bar{B}_t\Phi^\sigma(x_t,o_t)\gamma_t+v_t\label{eq:tv1}\\
    v_t&\sim\mathcal{N}(0,\Sigma_v)\label{eq:tv1}
\end{align}
% \vskip -0.1 true in
\end{subequations}}}
The mode $\sigma$ and weights $\gamma_t$ are unknown, although the feature matrices {\small{$\Phi^j(x_t,o_t)$}} are known {\small{$\forall j\in\{1,\dots,M\}$}}. The weights also evolve dynamically as
{\small{
\begin{subequations}\label{eq:gam}
% \vskip -0.25 true in
\begin{align}
    \gamma_{t+1}&=\gamma_t+n_t\label{eq:gam1}\\
    n_t&\sim\mathcal{N}(0,\Sigma_n)\label{eq:gam2}.
\end{align}
% \vskip -0.1 true in
\end{subequations}}}
The probability of mode $j$ at time $t$ is denoted as {\small{$p^j_t$}}, and obtained in a Bayesian fashion \cite{keener2010theoretical} using measurements {\small{$(x_t,o_t,x_{t-1},u_{t-1},o_{t-1})$}} as
{
% \vskip -0.25 true in
\small{
\begin{align}\label{eq:pjt}
    p^j_{t}=\frac{P[(x_t,o_t)|(x_{t-1},u_{t-1},o_{t-1}),\sigma=j]p^j_{t-1}}{\sum_{i=1}^MP[(x_t,o_t)|(x_{t-1},u_{t-1},o_{t-1}),\sigma=i]p^i_{t-1}}
\end{align}}
% \vskip -0.1 true in
}
with the initialization {\small{$p^j_0=\frac{1}{M},~\forall j\in\{1,\dots,M\}$}}, and likelihood {\small{$P[(x_t,o_t)|(x_{t-1},u_{t-1},o_{t-1}), \sigma=i]$}} given by the probability density function (pdf) of the random variable {\small{$(x_t,o_t)|\{(x_{t-1},u_{t-1},o_{t-1}), \sigma=i\}$}}.
\subsection{Constraints}
The traffic rules, speed, actuation constraints are given by chance constraints {\small{$\mathbb{P}\left[\left(x_{t+1},u_t\right)\not\in\mathcal{XU}_{t}\right]\leq\epsilon$}}, and the collision avoidance chance constraints are given by {\small{$\mathbb{P}\left[\left(x_t,o_t\right)\not\in\mathcal{C}_t\right]\leq\epsilon$}} for some given risk level {\small{$\epsilon>0$}},  where {\small{\begin{align*}\mathcal{XU}_{t}=&\{\left(x,u\right):f^{i\top}_{x,t}x+f^{i\top}_{u,t}u\le f^i_t, \forall i=1,..,n_{xu}\}\\
\mathcal{C}_t=&\{\left(x,o\right):g^{i\top}_{x,t}x+g^{i\top}_{o,t}o\leq g^i_t, \forall i=1,..,n_c\}.
\end{align*}}}
\subsection{SMPC Algorithm}
The SMPC optimization problem is given as follows
\small\begin{subequations}\label{opt:MPC_skeleton}
% \vskip -0.2 true in
\begin{align}
 \min_{\substack{\boldsymbol{\theta}_{t}}}&\quad \sum_{j=1}^Mp^j_t\cdot\mathbb{E}[\sum_{k=t}^{t+N-1}c_t(x^j_{k+1|t},u^j_{k|t})]\label{opt:obj}\\
 \text{s.t. }&~x^j_{k+1|t}=A_{k}x^j_{k|t}+B_{k} u^j_{k|t}+w_{k|t},\label{opt:EV_dyn}\\
&~o^j_{k+1|t}=\bar{A}_ko^j_{k|t}+\bar{B}_k\Phi^j(x^j_{k|t},o^j_{k|t})\gamma^j_{k|t}+v_{k|t},\label{opt:TV_dyn}\\
&~\gamma^j_{k+1|t}=\gamma^j_{k|t}+n_{k|t},\label{opt:gam_dyn}\\
&~u^j_{k|t}=\pi_{\theta_{k|t}}(x^j_{t|t},..,x^j_{k|t},o^j_{t|t},.., o^j_{k|t}),\label{opt:gen_pol_class}\\
&~\mathbb{P}\left[\left(x^j_{k+1|t},u^j_{k|t}\right)\not\in\mathcal{XU}_{k}\right]\leq\epsilon\label{opt:xu_constr}\\
&~\mathbb{P}\left[\left(x^j_{k+1|t},o^j_{k+1|t}\right)\not\in\mathcal{C}_{k+1}\right]\leq\epsilon\label{opt:ca_constr}\\
&~x^j_{t|t}=x_t,\ o^j_{t|t}=o_t,\label{opt:init}\\
&~\gamma^j_{t|t}\sim\mathcal{N}(\hat{\gamma}^j_{t|t},\Sigma^j_{t|t})\label{opt:gam_init}\\
&~\forall j=1,.., M, \forall k=t,..,t+N-1\nonumber
\end{align}
% \vskip -0.1 true in
\end{subequations}\normalsize
where $c_t(\cdot)$ is the stage cost, notation $q_{k|t}^j$ denotes the prediction of quantity $q$ for time $k$ and mode $j$, at time $t$. Problem \eqref{opt:MPC_skeleton} is solved to find the optimal decision variables {\small{$\boldsymbol{\theta}^*_t=(\theta_{t|t}^*,..,\theta_{t+N-1|t}^*)$}}  parameterizing the control laws in \eqref{opt:gen_pol_class}. The control at time $t$ is {\small{$u_t=\pi_{\theta_{t|t}^\ast}\left(x_t,o_t\right)$}} and \eqref{opt:MPC_skeleton} is solved again at time $t+1$. Since the mode $\sigma$ in \eqref{eq:tv1} is unobservable, the predictions for all modes (in \eqref{opt:EV_dyn},\eqref{opt:TV_dyn}) are initialized using the observed $x_t,o_t$ (via \eqref{opt:init}), and the control laws {\small{$\pi_{\theta_{k|t}}\left(\cdot\right)$}} are mode-agnostic. Problem \eqref{opt:MPC_skeleton} also requires initial distributions of the weights in \eqref{opt:gam_init} which are obtained using the Kalman Filter (KF) for \eqref{eq:gam}, with \eqref{eq:tv_dyn} being the observation model.

The goal is to design {\small{$\pi_{\theta_{k|t}}\left(\cdot\right)$}}, {\small{$\forall k=t,..,t+N-1$}}, such that solving \eqref{opt:MPC_skeleton} is computationally tractable. The control laws in \eqref{opt:gen_pol_class} offer greater flexibility for satisfying \eqref{opt:xu_constr},\eqref{opt:ca_constr} than using a single sequence of controls {\small{$\pi_{\theta_{k|t}}\left(\cdot\right)\equiv\theta_{k|t}$}} (which are independent of different EV, TV trajectory realizations). To further reduce conservatism, the control laws must also serve a dual role of predicting and estimating the distributions of $\gamma_{k|t}^j$ using EV, TV trajectory realizations {\small{$x_{t|t}^j,..,x_{k|t}^j,o_{t|t}^j,..,o_{k|t}^j$}} along the prediction horizon.

\section{SMPC FOR MULTI-MODAL, INTERACTION-AWARE PREDICTIONS}\label{sec:SMPC}

The details of our SMPC design are as follows.
\subsection{Linearized Predictions}\label{ssec:lin_pred}
\textbf{Linearization of (5c):} The nonlinear term {\small{$\Phi^j(x^j_{k|t},o^j_{k|t})\gamma^j_{k|t}$}} in \eqref{opt:TV_dyn} is linearized about {\small{$(\bar{x}^j_{k+1|t-1},\bar{o}^j_{k+1|t-1} \hat{\gamma}^j_{t|t})$}} using the previous solution of \eqref{opt:MPC_skeleton}, given as {\small{
% \vskip -0.2 true in
\begin{align*}
\bar{x}^j_{k+1|t-1}&=A_k\bar{x}^j_{k|t-1}+B_k\bar{u}^j_{k|t-1},\\\bar{o}^j_{k+1|t-1}&=\bar{A}_k\bar{o}^j_{k|t-1}+\bar{B}_k\Phi^j(\bar{x}^j_{k|t-1},\bar{o}^j_{k|t-1})\hat{\gamma}^j_{t|t},\\ \bar{u}^j_{k|t-1}&=\pi_{\theta^*_{k|t-1}}(x_{t-1},..,\bar{x}^j_{k|t},o_{t-1},..,\bar{o}^j_{k|t-1}),\\ \bar{x}^j_{t|t-1}&=x_t,\ \bar{o}^j_{t|t-1}=o_t.
% \vskip -0.2 true in
\end{align*}
% \vskip -0.1 true in
}} The linearized term is denoted as
{\small{
$\bar{B}_k\Phi_j(x^j_{k|t},o^j_{k|t})\gamma^j_{k|t}\approx G^j_{k|t}\gamma^j_{k|t}+P^j_{k|t}x^j_{k|t}+Q^j_{k|t}o^j_{k|t}+l^j_{k|t}
$}}, where the coefficient matrices are given by
{\small{
% \vskip -0.2 true in
\begin{align*}G^j_{k|t}&=\bar{B}_k\Phi^j(\bar{x}^j_{k+1|t-1},\bar{o}^j_{k+1|t-1}),\\ [P^j_{k|t}\ Q^j_{k|t}]&=\bar{B}_k\sum_{i=1}^{n_\gamma}\nabla\Phi_i^j(\bar{x}^j_{k+1|t-1},\bar{o}^j_{k+1|t-1})[\hat{\gamma}^j_{t|t}]_i,\\  l^j_{k|t}&=-\bar{B}_k(P^j_{k|t}\bar{x}^j_{k+1|t-1}+Q^j_{k|t}\bar{o}^j_{k+1|t-1}),
\end{align*}
% \vskip -0.1 true in
}} 
using notation {\small{$\Phi^j(x,o)=[\Phi^j_1(x,o)\ ..\ \Phi^j_{n_\gamma}(x,o)]$}}.

\textbf{Prediction and Estimation of Weights:}
Let {\small{$\tilde{\gamma}^j_{k|t}$}} denote the random variable {\small{$\gamma^j_{k|t}$}} conditioned on {\small{$x_{t|t}^j,..,x_{k|t}^j,o_{t|t}^j,..,o_{k|t}^j$}}, with {\small{$\tilde{\gamma}^j_{t|t}=\gamma^j_{t|t}$}}. Given a prior distribution {\small{$\tilde{\gamma}^j_{k-1|t}\sim\mathcal{N}(\hat{\gamma}^j_{k-1|t}, \Sigma^j_{k-1|t})$}}, we compute the posterior distribution using the measurements {\small{$(x_{k-1|t}^j,o_{k-1|t}^j,o_{k|t}^j)$}} and \eqref{opt:TV_dyn}  to construct a measurement model for {\small{$\tilde{\gamma}^j_{k-1|t}$}} as
{\small{
% \vskip -0.2 true in
\begin{align}\label{eq:gamm_meas_model}
    &y^j_{k|t}:=o^j_{k|t}-(\bar{A}_{k-1}+Q^j_{k-1|t})o^j_{k-1|t}-P^j_{k-1|t}x^j_{k-1|t}-l^j_{k-1|t},\nonumber\\
    &y^j_{k|t}=G^j_{k-1|t}\gamma^j_{k-1|t}+v_{k-1|t},
\end{align}
% \vskip -0.1 true in
}}
where we have used the TVs' state measurements and linearised dynamics to define the output in the first equation. The distribution of {\small{$\tilde{\gamma}^j_{k|t}$}} is obtained from the ``update" step of the KF (with \eqref{opt:gam_dyn} as the dynamics model and \eqref{eq:gamm_meas_model} as the measurement model) to give
{\small{
% \vskip -0.1 true in
\begin{align}\label{eq:gamm_update}
\tilde{\gamma}^j_{k|t}=(I-K^j_{k|t}G^j_{k-1|t})\tilde{\gamma}^j_{k-1|t}+K^j_{k|t}y^j_{k|t}+n_{k|t}\end{align}
% \vskip -0.1 true in
}}
where {\small{$K^j_{k|t}=\Sigma^j_{k-1|t}G^{j\top}_{k-1|t}(\Sigma_v+G^j_{k-1|t}\Sigma^j_{k-1|t}G^{j\top}_{k-1|t})^{-1}$}} is the Kalman gain. The mean and covariance are given by {\small{ \begin{align}\hat{\gamma}^j_{k|t}&=(I-K^j_{k|t}G^j_{k-1|t})\hat{\gamma}^j_{k-1|t}+K^j_{k|t}y^j_{k|t}\nonumber\\
\Sigma^j_{k|t}&=(I-K^j_{k|t}G^j_{k-1|t})\Sigma^j_{k-1|t}+\Sigma_n.\end{align}
}}In terms of the initial distribution in \eqref{opt:gam_init}, \eqref{eq:gamm_update} can be alternatively expressed as 
{\small{
\vskip -0.1 true in
\begin{align}\label{eq:gam_tilde_batch}
    \tilde{\gamma}^j_{k|t}=\prod_{i=t+1}^k W^j_{i|t}(\hat{\gamma}^j_{t|t}+n^j_{t})+\sum_{i=t+1}^{k}\prod_{l=i}^{k-1} W^j_{l|t}(K^j_{i|t}y^j_{i|t}+n_{i|t})
\end{align}
\vskip -0.1 true in
}}where {\small{$n^j_{t}\sim\mathcal{N}(0,\Sigma^j_{t|t})$}} and {\small{$ W^j_{k|t}=I-K^j_{k|t}G^j_{k-1|t}$}}. The initial distribution \eqref{opt:gam_init} is also similarly obtained by KF from the TVs' state measurements $o_{t},o_{t-1}$. 

\textbf{Consolidated TV Model:} Denote {\small{$\tilde{o}^j_{k+1|t}$}} as the random variable {\small{$o^j_{k+1|t}$}} conditioned on {\small{$x_{t|t}^j,..,x_{k|t}^j,o_{t|t}^j,..,o_{k|t}^j$}}, which is given by {\small{\begin{align*}
    \tilde{o}^j_{k+1|t}&=(\bar{A}_k+Q^j_{k|t})o^j_{k|t}+P^j_{k|t}x^j_{k|t}+G^j_{k|t}\tilde{\gamma}^j_{k|t}+l^j_{k|t}+v_{k|t}
\end{align*}}}
Using \eqref{eq:gam_tilde_batch} and defining a new random variable {\small{$z^j_{k|t}$}}, as {\vskip -0.1 true in\small{\begin{align}\label{eq:z_def}z^j_{k|t}=v_{k|t}+G^j_{k|t}(\prod_{i=t+1}^k W^j_{i|t}n^j_{t}+\sum_{i=t+1}^{k}\prod_{l=i}^{k-1} W^j_{l|t}n_{i|t}),\end{align}}\vskip -0.1 true in} we can replace \eqref{opt:TV_dyn} and \eqref{opt:gam_dyn} by the following consolidated model 
{
% \vskip -0.2 true in
\small{
\begin{align}\label{eq:tv_cnsldtd}
    \tilde{o}^j_{k+1|t}&=(\bar{A}_k+Q^j_{k|t})o^j_{k|t}+P^j_{k|t}x^j_{k|t}+G^j_{k|t}\hat{\gamma}^j_{k|t}+z^j_{k|t}+l^j_{k|t}\nonumber\\
    \hat{\gamma}^j_{k|t}&=\prod_{i=t+1}^k W^j_{i|t}\hat{\gamma}^j_{t|t}+\sum_{i=t+1}^{k}\prod_{l=i}^{k-1} W^j_{l|t}K^j_{i|t}y^j_{i|t}.
\end{align}}
% \vskip -0.14 true in
}
Note that $z^j_{k|t}$ serves as the ``effective" process noise for TV prediction model, which can be measured using measurements  of {\small{$o_t,x_t,..,o_{k|t}^j, x_{k|t}^j, o_{k+1|t}^j$}}. In contrast, see that {\small{$v_{k|t}$}} can't be measured using \eqref{opt:TV_dyn} because $\gamma^j_{k|t}$ is unknown. This observation will be important for designing the control laws {\small{$\pi_{\theta_{k|t}}\left(\cdot\right)$}}, {\small{$\forall k=t,..,t+N-1$}}. 

\textbf{Stacked Predictions:} Let {\small{$\mathbf{x}_t^j=[x_{t},..,x^j_{t+N|t}]$}} (similarly for {\small{$\tilde{\mathbf{o}}_t^j$}}), {\small{$\tilde{\mathbf{o}}_t^j=[\tilde{o}^j_{t+1|t},..,\tilde{o}^j_{t+N|t}]$}}, {\small{$\mathbf{u}_t^j=[u^j_{t|t},..,u^j_{t+N-1|t}]$}} (similarly for {\small{$\mathbf{w}_t,\mathbf{n}_t, \hat{\boldsymbol{\gamma}}^j_t, \boldsymbol{\gamma}^j_t,  \mathbf{z}^j_t, \mathbf{l}^j_t$}}). The stacked state predictions for the EV can be expressed as a function of the current state {\small{$x_t$}} and stacked inputs {\small{$\mathbf{u}_t^j$}} as
{\small{\begin{align}\label{eq:x_stacked}\mathbf{x}_t^j&=\mathbf{A}_tx_t+\mathbf{B}_t\mathbf{u}^j_t+\mathbf{E}_t\mathbf{w}_t.\end{align}}}  For the TVs, the stacked,conditioned predictions {\small{$\tilde{\mathbf{o}}_t^j$}} are given by
{\small{\begin{align*}\tilde{\mathbf{o}}_t^j&=\tilde{\mathbf{A}}^j_t\mathbf{o}^j_t+\tilde{\mathbf{P}}^j_t\mathbf{x}^j_t+\tilde{\mathbf{G}}^j_t\hat{\boldsymbol{\gamma}}^j_t+\mathbf{z}^j_t+\mathbf{l}^j_t,\\ \mathbf{o}_t^j&=\bar{\mathbf{A}}^j_to_t+\mathbf{P}^j_t\mathbf{x}^j_t+\mathbf{G}^j_t\boldsymbol{\gamma}^j_t+\mathbf{F}^j_t(\mathbf{v}_t+\mathbf{l}^j_t).\end{align*}}}Using {\small{$\boldsymbol{\gamma}^j_t=\mathbf{G}\hat{\gamma}_{t|t}+\mathbf{\Gamma}\mathbf{n}^j_t$}} (from \eqref{opt:gam_dyn},\eqref{opt:gam_init}) and {\small{$\hat{\boldsymbol{\gamma}}^j_t=\mathbf{\Gamma}^{j,\gamma}_t\hat{\gamma}_{t|t}+\mathbf{\Gamma}^{j,x}_t\mathbf{x}^j_t+\mathbf{\Gamma}^{j,o}_t\mathbf{o}^j_t+\mathbf{\Gamma}^{j,l}_t\mathbf{l}^j_t$}} (from \eqref{eq:gamm_meas_model}, \eqref{eq:tv_cnsldtd}) where {\small{$\mathbf{n}^j_t=[n^j_t, \mathbf{n}_t]$}}, the stacked, conditioned predictions {\small{$\tilde{\mathbf{o}}_t^j$}} are explicitly given as  {\small{\begin{align}\label{eq:o_stacked}\tilde{\mathbf{o}}_t^j=&\bar{\bar{\mathbf{A}}}^{j,o}_to_t+\bar{\bar{\mathbf{A}}}^{j,x}_tx_t+\bar{\bar{\mathbf{B}}}^{j}_t\mathbf{u}^j_t+\bar{\bar{\mathbf{G}}}^j_t\hat{\gamma}_{t|t}+\bar{\bar{\mathbf{F}}}^{j,n}_t\mathbf{n}^j_t+\mathbf{z}^j_t\nonumber\\&+\bar{\bar{\mathbf{F}}}^{j,v}_t\mathbf{v}_t+\bar{\bar{\mathbf{F}}}^{j,w}_t\mathbf{w}_t+\bar{\bar{\mathbf{L}}}^j_t.\end{align}}} 
We defer the various matrix definitions to the appendix.

\subsection{Control Law Parameterization}
We use the affine disturbance feedback parameterization for our control law,
{
\vskip -0.2 true in
\small{\begin{align}\label{eq:pol}
    &\pi_{\theta_{k|t}}(x_{t},..,x^j_{k|t},o_{t},.., o^j_{k|t})=h_{k|t}+\sum_{i=t}^{k-1}M^w_{i,k|t}w_{i|t}+M^z_{i,k|t}z^j_{i|t}
\end{align}}
\vskip -0.1 true in
}
which is a function of the past process noise realizations {\small{$w_{t|t}, z^j_{t|t},..,w_{k-1|t}, z^j_{k-1|t}$}}, 
with mode-agnostic parameters {\small{$\theta_{k|t}=(h_{k|t},\{M^w_{i,k|t}, M^z_{i,k|t}\}_{i=t}^{k-1})$}} (see \cite{goulart2006optimization} for equivalence of disturbance feedback and state feedback). For an intuitive explanation of the equivalence, notice that each {\small{$w_{i|t}$}} can be measured from consecutive EV state measurements {\small{$x^j_{i|t}, x^j_{i+1|t}$}} using \eqref{opt:EV_dyn}  and each {\small{$z^j_{i|t}$}} can be measured from {\small{$o_t,x_t,..,o_{i|t}^j, x_{i|t}^j, o_{i+1|t}^j$}} using \eqref{eq:tv_cnsldtd}. Conversely, given the process noise sequence {\small{$w_{t|t}, z^j_{t|t},..,w_{k-1|t}, z^j_{k-1|t}$}}, the EV and TV state trajectories {\small{$o_t,x_t,..,o_{k|t}^j, x_{k|t}^j$}} are completely determined. Thus, there exists an invertible transformation {\small{$\mathcal{T}(\cdot)$}} such that {\small{$o_t,x_t,..,o_{k|t}^j, x_{k|t}^j=\mathcal{T}(w_{t|t}, z^j_{t|t},..,w_{k-1|t}, z^j_{k-1|t})$}}.

The stacked control inputs are given by {\small{\begin{align}\label{eq:u_stacked}\mathbf{u}^j_t=\mathbf{h}_t+\mathbf{M}^w_t\mathbf{w}_t+\mathbf{M}^z_t\mathbf{z}^j_t\end{align}}}where the stacked control parameter matrices {\small{$\mathbf{h}_t, \mathbf{M}_t^w, \mathbf{M}_t^z$}} are defined in the appendix. Note that these parameters will be the decision variables for our SMPC, and thus optimized online.
\subsection{Chance Constraint Reformulation} 
The chance constraints \eqref{opt:xu_constr},\eqref{opt:ca_constr} are reformulated using the following result from \cite{calafiore2006distributionally}: {\small{\begin{align*}&\mathbb{P}(a^\top x >b)\leq \epsilon, x\sim\mathcal{N}(\mu,\Sigma)\\\Leftrightarrow& a^\top\mu+\mathfrak{Q}(1-\epsilon)||\sqrt{\Sigma}a ||_2\leq b\end{align*}}} where {\small{$\mathfrak{Q}(\cdot)$}} is the quantile function of {\small{$\mathcal{N}(0,1)$}}. The distributions of the stacked predictions {\small{$\mathbf{x}^j_t, \tilde{\mathbf{o}}^j_t  $}} \eqref{eq:x_stacked}, \eqref{eq:o_stacked} and control laws {\small{$\mathbf{u}^j_t $}} \eqref{eq:u_stacked} are determined by the random variables {\small{$\mathbf{w}_t, \mathbf{v}_t, \mathbf{n}^j_t $}}. Note that {\small{$\mathbf{z}^j_t=\mathbf{v}_t+\mathbf{\Gamma}^{j,z}_t\mathbf{n}^j_t$}}, where {\small{$\mathbf{\Gamma}^{j,z}_t $}} is defined in the appendix. Define constant matrices {\small{$S^x_k, S^o_k, S^u_k$}} such that {\small{$S^x_k\mathbf{x}^j_t=x^j_{k|t}, S^o_k\tilde{\mathbf{o}}^j_t=\tilde{o}^j_{k|t}, S^u_k\mathbf{u}^j_t=u^j_{k|t}$}}, to recover predictions at the $k^{th}$ time step from the stacked predictions \eqref{eq:x_stacked}, \eqref{eq:o_stacked}, \eqref{eq:u_stacked}. Then the state-input chance constraints \eqref{opt:xu_constr}, and the collision avoidance chance constraints \eqref{opt:ca_constr} can be rewritten as
{\footnotesize{\begingroup\allowdisplaybreaks
\begin{subequations}
\begin{align}
    &\mathbb{P}((x^j_{k+1|t},u^j_{k|t})\not\in\mathcal{XU}_k)\leq\epsilon\nonumber\\
    &\Leftrightarrow\mathbb{P}(f^{i\top}_{x,k}x^j_{k+1|t}+f^{i\top}_{u,k}u^j_{k|t}>f^i_k,~\forall i=1,..,n_{xu})\leq\epsilon\nonumber\\
    &\Leftarrow\mathbb{P}(f^{i\top}_{x,k}x^j_{k+1|t}+f^{i\top}_{u,k}u^j_{k|t}>f^i_k)\leq\frac{\epsilon}{n_{xu}},~\forall i=1,..,n_{xu}\nonumber\\
    &\Leftrightarrow f^{i\top}_{x,k}S^x_{k+1}(\mathbf{A}_tx_t+\mathbf{B}_t\mathbf{h}_t)+f^{i\top}_{u,t}S^u_{k}\mathbf{h}_t-f^i_k\leq-\mathfrak{Q}(1-\frac{\epsilon}{n_{xu}})\times\nonumber\\
    &\left\Vert\sqrt{\mathbf{\Sigma}}^{j}_t\begin{bmatrix}(f^{i\top}_{x,k}S^x_{k+1}(\mathbf{B}_t\mathbf{M}^w_t+\mathbf{E}_t)+f^{i\top}_{u,k}S^u_{k}\mathbf{M}^w_t)^\top\\ (f^{i\top}_{x,k}S^x_{k+1}\mathbf{B}_t\mathbf{M}^z_t+f^{i\top}_{u,k}S^u_{k}\mathbf{M}^z_t)^\top\\\mathbf{\Gamma}^{j,z\top}_t(f^{i\top}_{x,k}S^x_{k+1}\mathbf{B}_t\mathbf{M}^z_t+f^{i\top}_{u,k}S^u_{k}\mathbf{M}^z_t)^\top\end{bmatrix} \right\Vert_2,\nonumber\\&~~~~~~~~~~~~~~~ \forall i=1,..,n_{xu}\label{eq:const_xu}\\
    &\mathbb{P}((x^j_{k|t},o^j_{k|t})\not\in\mathcal{C}_k)\leq\epsilon\nonumber\\
    &\Leftrightarrow\mathbb{P}(g^{i\top}_{x,k}x^j_{k|t}+g^{i\top}_{o,k}o^j_{k|t}>g^i_t,~\forall i=1,..,n_{c})\leq\epsilon\nonumber\\
    &\Leftarrow\mathbb{P}(g^{i\top}_{x,k}x^j_{k|t}+g^{i\top}_{o,k}o^j_{k|t}>g^i_t)\leq\frac{\epsilon}{n_{c}},~\forall i=1,..,n_{c}\nonumber\\
    &\Leftrightarrow g^{i\top}_{x,k}S^x_{k}(\mathbf{A}_tx_t+\mathbf{B}_t\mathbf{h}_t)+g^{i\top}_{o,k}S^o_{k}(\bar{\bar{\mathbf{A}}}^{j,o}_to_t+\bar{\bar{\mathbf{A}}}^{j,x}_tx_t\nonumber\\
    &+\bar{\bar{\mathbf{B}}}^{j}_t\mathbf{h}_t+\bar{\bar{\mathbf{G}}}^j_t\hat{\gamma}_{t|t}+\bar{\bar{\mathbf{L}}}^j_t)-g^i_k\leq-\mathfrak{Q}(1-\frac{\epsilon}{n_{c}})\times\nonumber\\
    &\left\Vert\sqrt{\mathbf{\Sigma}}^j_t\begin{bmatrix}(g^{i\top}_{x,k}S^x_{k}(\mathbf{B}_t\mathbf{M}^w_t+\mathbf{E}_t)+g^{i\top}_{o,k}S^o_{k}(\bar{\bar{\mathbf{B}}}^j_t\mathbf{M}^w_t+\bar{\bar{\mathbf{F}}}^{j,w}_t))^\top\\ (g^{i\top}_{x,k}S^x_{k}\mathbf{B}_t\mathbf{M}^z_t+g^{i\top}_{o,k}S^o_{k}(\bar{\bar{\mathbf{B}}}^{j}_t\mathbf{M}^z_t+I+\bar{\bar{\mathbf{F}}}^{j,v}_t))^\top\\(g^{i\top}_{x,k}S^x_{k}\mathbf{B}_t\mathbf{M}^z_t\mathbf{\Gamma}^{j,z}_t+g^{i\top}_{o,k}S^o_{k}((\bar{\bar{\mathbf{B}}}^{j}_t\mathbf{M}^z_t+I)\mathbf{\Gamma}^{j,z}_t+\bar{\bar{\mathbf{F}}}^{j,n}_t))^\top\end{bmatrix} \right\Vert_2\nonumber\\
    &~~~~~~~~~~~~~~~\forall i=1,..,n_{c}\label{eq:const_ca}
\end{align}
\end{subequations}
\endgroup}}
where {\small{$\mathbf{\Sigma}^j_t=\text{blkdiag}(\mathbf{\Sigma}_w, \mathbf{\Sigma}_v,\Sigma^j_{t|t},\mathbf{\Sigma}_n), \mathbf{\Sigma}_w=I_N\otimes\Sigma_w$}} (similarly for {\small{$\mathbf{\Sigma}_v, \mathbf{\Sigma}_n$}}). Note that in the second step of each reformulation, we use Boole's inequality to obtain individual chance constraints whose satisfaction imply satisfaction of the original joint chance constraints. These constraint reformulations are of the form {\small{$\mathfrak{Q}(1-\tilde{\epsilon})\Vert C\boldsymbol{\theta}_t+d \Vert_2\leq a^\top\boldsymbol{\theta}_t+b $}}, where {\small{$\boldsymbol{\theta}_t=(\mathbf{h}_t, \mathbf{M}^w_t, \mathbf{M}^z_t)$}} are the control laws' parameters and {\small{$\tilde{\epsilon}=\dfrac{\epsilon}{n_{xu}}$}} for \eqref{eq:const_xu}, {\small{$\tilde{\epsilon}=\dfrac{\epsilon}{n_{c}}$}} for \eqref{eq:const_ca}. These are convex, second-order cone constraints in the control laws' parameters {\small{$\boldsymbol{\theta}_t=(\mathbf{h}_t, \mathbf{M}^w_t, \mathbf{M}^z_t)$}} when {\small{$\epsilon\leq\min\{\frac{n_{xu}}{2},\frac{n_{c}}{2}\}$}} \cite{boyd2004convex}.
\subsection{Cost Definition}
\textbf{Estimation of Mode Probabilities:} The trajectory cost for each mode is weighted by the mode probability, {\small{$p^j_t$}}, to prioritize minimizing the trajectory cost for more likely modes. For estimating {\small{$p^j_t$}} using \eqref{eq:pjt}, the likelihood functions are computed using the pdf which is obtained using the consolidated TV model \eqref{eq:tv_cnsldtd} as {\small{\begin{align}&P[(x_t,o_t)|(x_{t-1},u_{t-1},o_{t-1}), \sigma=i]\nonumber\\&=P[z^i_{t-1|t-1}=o_t-\bar{A}_{t-1}o_{t-1}-\bar{B}_{t-1}\Phi^i(x_{t-1},o_{t-1})\hat{\gamma}^i_{t-1|t-1}]\end{align}}} where {\small{$z^i_{t-1|t-1}\sim\mathcal{N}(0,\Sigma_v+G^i_{t-1|t-1}\Sigma^i_{t-1|t-1}G^{i\top}_{t-1|t-1})$}\vskip -0.1 true in}.

\textbf{Stage Cost:} We choose a convex, quadratic cost to penalise deviations from a given reference, {\small{$c_t(x,u)=(x-x^{\text{ref}}_{t+1})^\top C_x(x-x^{\text{ref}}_{t+1})+(u-u^{\text{ref}}_{t})^\top C_u(u-u^{\text{ref}}_{t}) $}} for positive definite {\small{$C_x,C_u$}}.

\textbf{Expected Trajectory Cost per Mode:} The expected trajectory cost for mode $j$  in \eqref{opt:obj} can be calculated as 
{
\vskip -0.1 true in
\small{\begin{align}\label{eq:cost}
    &\mathbb{E}[\sum_{k=t}^{t+N-1}c(x^j_{k+1|t},u^j_{k|t})]\nonumber\\
    &=\mathbb{E}[(\mathbf{x}^j_{t}-\mathbf{x}^{\text{ref}}_t)^\top\mathbf{C}_x(\mathbf{x}^j_{t}-\mathbf{x}^{\text{ref}}_t)+(\mathbf{u}^j_{t}-\mathbf{u}^{\text{ref}}_t)^\top\mathbf{C}_u(\mathbf{u}^j_{t}-\mathbf{u}^{\text{ref}}_t)]\nonumber\\    &=\mathbf{h}_t^\top(\mathbf{B}^\top_t\mathbf{C}_x\mathbf{B}_t+\mathbf{C}_u)\mathbf{h}_t\nonumber\\		&+\text{tr}\big((\mathbf{B}_t^\top\mathbf{C}_x\mathbf{B}_t+\mathbf{C}_u)[\mathbf{M}_t^w\ \mathbf{M}_t^z\ \mathbf{M}_t^z\mathbf{\Gamma}^{j,z}_t] \mathbf{\Sigma}^j_t[\mathbf{M}_t^w\ \mathbf{M}_t^z\ \mathbf{M}_t^z\mathbf{\Gamma}^{j,z}_t]^\top\nonumber\\
    &+(\mathbf{A}_tx_{t}-\mathbf{x}^{\text{ref}}_t)^\top\mathbf{C}_x(\mathbf{A}_tx_{t}-\mathbf{x}^{\text{ref}}_t+2\mathbf{B}_t\mathbf{h}_t)\nonumber\\
	&-2\mathbf{u}^{\text{ref}\top}_t\mathbf{C}_u\mathbf{B}_t\mathbf{h}_t+\text{tr}(\mathbf{C}_x\mathbf{E}_t\boldsymbol{\Sigma}_w\mathbf{E}^\top_t)
\end{align}}
% \vskip -0.1 true in
}
where {\small{$\mathbf{C}_x=I_N\otimes C_x, \mathbf{C}_u=I_N\otimes C_u,\ \mathbf{x}^{\text{ref}}_t=[x^{\text{ref}}_{t+1},..,x^{\text{ref}}_{t+N}], \mathbf{u}^{\text{ref}}_t=[u^{\text{ref}}_{t},..,u^{\text{ref}}_{t+N-1}] $}} . Due to the positive definiteness of {\small{$C_x,C_u$}}, the cost is convex and quadratic in the control laws' parameters {\small{$\boldsymbol{\theta}_t=(\mathbf{h}_t, \mathbf{M}^w_t, \mathbf{M}^z_t)$}}.
\subsection{SMPC Optimization Problem}
The SMPC optimization problem is given in batch form, by explicitly substituting for the predictions \eqref{opt:EV_dyn},\eqref{eq:tv_cnsldtd} and parameterised control laws \eqref{eq:pol}, to yield 
{\vskip -0.2 true in
\small{
\begin{align}\label{opt:SMPC}
 \min_{\substack{\mathbf{h}_{t},\mathbf{M}^w_t, \mathbf{M}^z_t}}&\quad \sum_{j=1}^Mp^j_t\cdot(\text{Cost per mode, eq. \eqref{eq:cost})}\nonumber\\
 \text{s.t. }&~\text{State-input constraints, eq. \eqref{eq:const_xu}}\nonumber\\
&~\text{Collision avoidance constraints, eq. \eqref{eq:const_ca}}\nonumber\\
&~\forall j=1,.., M, \forall k=t,..,t+N-1
\end{align}}
\vskip -0.1 true in}
The optimization problem \eqref{opt:SMPC} is a convex, second-order cone programming problem that can be solved online for the control input {\small{$u_t=h^*_{t|t}$}}.
\section{SIMULATIONS}\label{sec:sim}
\subsection{Problem Formulation for Longitudinal Control:} 
\begin{figure}[!h]
    \vskip -0.1 true in
    \centering
    \includegraphics[scale=0.7]{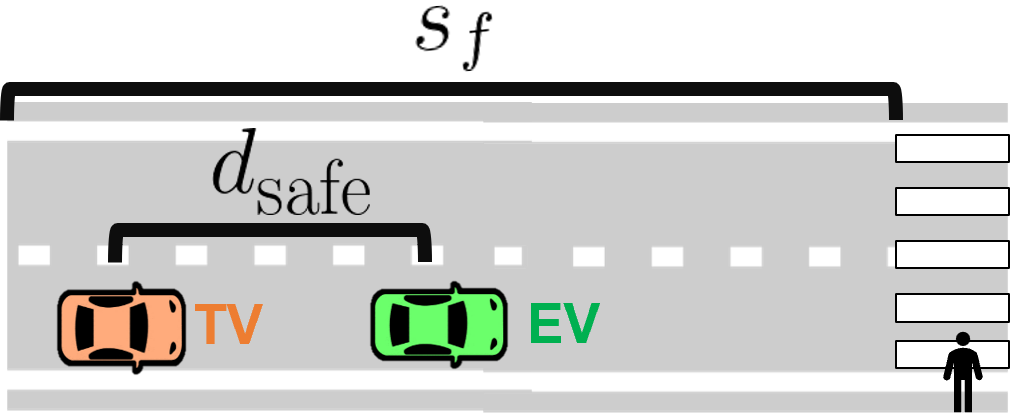}
    \vskip -0.1 true in
    \caption{EV must stop for the pedestrian while keeping safe distance from TV.}\label{fig:longeg}
    \vskip -0.1 true in
\end{figure}

Consider the longitudinal control problem as depicted in fig.~\ref{fig:longeg}, described by point-mass models with longitudinal position, speed describing the states, acceleration as the control input, for both the EV and TV respectively denoted as {\small{$x_t=[s_t,v_t], u_t=a_t,  o_t=[s^{o}_t, v^{o}_t]$}}. The system matrices are time-invariant and obtained from Euler-discretization as
{\small{\begin{align*}A_t=\bar{A}_t=\begin{bmatrix}1&\text{dt}\\0&0\end{bmatrix}, B_t=\bar{B}_t=\begin{bmatrix}0\\\text{dt}\end{bmatrix}.\end{align*}}\vskip -0.1 true in}

The TV control is assumed to have {\small{$M=2$}} modes with {\small{$n_\gamma=1$}} feature each, given by
{
% \vskip -0.2 true in
\small{\begin{subequations}
\begin{align}
\Phi^1(x_t,o_t)&=k^1_s(s_f-d_{\text{safe}}-s^o_t)+k^1_v(0-v^o_t)\\
\Phi^2(x_t,o_t)&=k^2_s(s_t-d_{\text{safe}}-s^o_t)+k^2_v(v_t-v^o_t)
\end{align}
\end{subequations}}
% \vskip -0.1 true in
}
where mode 1 is an EV-agnostic PD control to stop at {\small{$s_f-d_{\text{safe}}$}}, and mode 2 is a PD control to follow behind the EV by {\small{$d_{\text{safe}}$}}. The car-following control for mode 2 is adapted from \cite{milanes2014modeling}, with constant headway $d_{\text{safe}}$. Thus, the TV's control law is given by
{\small{\begin{align*}
u^o(x_t,o_t,\gamma_t, \sigma)=\begin{cases}\gamma_t\Phi^1(x_t,o_t), & \text{if } \sigma=1\\
\gamma_t\Phi^2(x_t,o_t), & \text{if } \sigma=2
\end{cases},
\end{align*}}}
where the mode $\sigma$, and the weight $\gamma_t$ are unknown to the EV, and thus need to be inferred online. The challenge is that the EV must stop for the pedestrian, and reach the stop fast enough to avoid the TV if {\small{$\sigma=1$}}, but must slow down smoothly to guide the TV to stop if {\small{$\sigma=2$}}, to prevent colliding with the TV.

The sets describing the state-input constraints and collision avoidance constraints are given as {\small{\begin{align*}\mathcal{XU}_t&=\{s_{t+1}\leq s_f, 0\leq v_{t+1}\leq v_{\text{max}}, a_{\text{min}}\leq a_t\leq a_{\text{max}}\}\\
\mathcal{C}_t&=\{s_{t}-s^o_t\geq d_{\text{safe}}\}.\end{align*} }} The cost is chosen to penalise deviations from the set-point {\small{$x^{\text{ref}}=[s_f,0], u^{\text{ref}}=0$}} and the EV acceleration $a_t$ is obtained by solving \eqref{opt:SMPC}.
\subsection{Simulation Results:}
The various parameters used in our simulation are given in Table~\ref{tab:sim_param}.
\begin{table}[!ht]
\centering
\caption{Parameters in SMPC simulations}\label{tab:sim_param}
\vskip -0.1 true in
\begin{tabular}{|| c | c ||} 
 \hline
 $\text{dt} = 0.1 s$ & $v_{\text{max}}=14 ms^{-1}$\\ 
 \hline
 $d_{\text{safe}}= 7 m$ & $v_{\text{min}}= 0 ms^{-1}$ \\
 \hline
 $s_f= 50 m$ & $a_{\text{max}}=3.5 ms^{-2}$\\
 \hline
  $N=12$ & $a_{\text{min}}= -6 ms^{-2}$\\
 \hline
 $\epsilon=0.1$ & $C_x=\text{diag}(50, 20)$\\
 \hline
 $\Sigma_w=\text{diag}(10^{-3},10^{-2})$ & $C_u=10$\\
 \hline
 $\Sigma_v=\text{diag}(10^{-2},10^{-1})$ & $k^1_s=1, k^1_v=6$\\
 \hline
 $\Sigma_n=0.5$ & $k^2_s=10^{-2}, k^1_v=1$\\
 \hline
\end{tabular}
\vskip -0.1 true in
\end{table}
We demonstrate the performance of the proposed SMPC design in closed-loop, for both {\small{$\sigma=1,\sigma=2$}} scenarios. For each scenario, we initialise the simulation with EV and TV states $x_0=[0,11]$, $o_0=[-9,15]$, and weight estimates $\hat{\gamma}^1_{0|0}=\hat{\gamma}^2_{0|0}=0$, $\Sigma^1_{0|0}=\Sigma^2_{0|0}=1$.
\begin{figure}[!h]
    \vskip -0.1 true in
    \centering
    \includegraphics[width=1.02\columnwidth]{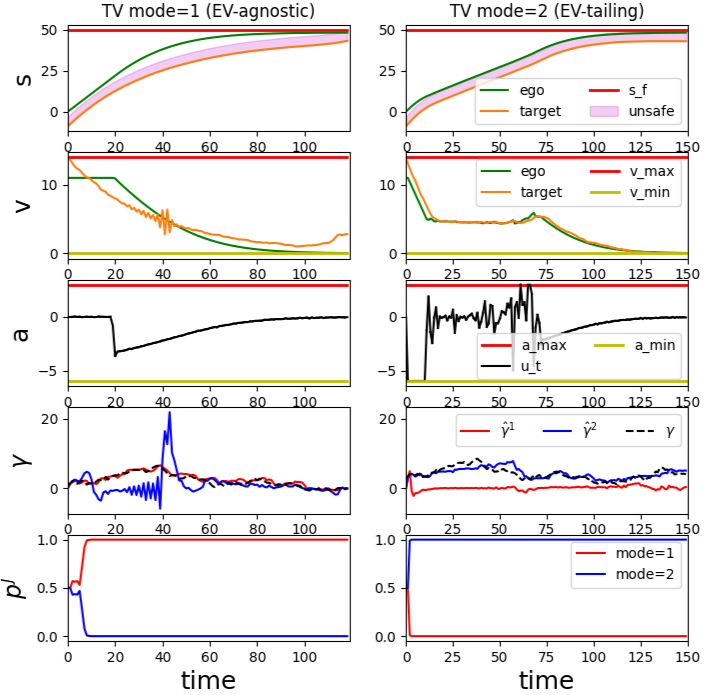}
    \vskip -0.1 true in
    \caption{The EV is able to stop (and also guide the TV to stop for {\small{$\sigma=2$}}) while satisfying constraints, tracking the unknown weight {\small{$\gamma_t$}} and identifying the correct mode {\small{$\sigma$}} with high probability.}\label{fig:sim}
    \vskip -0.1 true in
\end{figure}
\subsection*{Discussion:}
Fig.~\ref{fig:sim} presents the simulation results for both scenarios, side by side. In the longitudinal position plots, we see that the EV's longitudinal position remains outside the pink unsafe band, i.e., {\small{$s_t\geq s^o_t+d_{\text{safe}}$}}, for both scenarios. From the speed and acceleration plots, we also see that EV is able to slow down to a stop at {\small{$s_f$}}, while also guiding the TV to stop at a safe distance in the scenario $\sigma=2$. In the fourth row, we see that the proposed estimation strategy allows close tracking of the weight true $\gamma_t$ by $\hat{\gamma}^j_t$, for the true mode $j=\sigma$. Finally, the last row shows that the true mode $\sigma$ was identified with high probability in both scenarios using the Bayesian updates. 
\subsection{Ablation Study:}
We demonstrate the benefits of the including feedback in predictions of the SMPC design \eqref{opt:SMPC} via ablation studies. The SMPC \eqref{opt:SMPC} is compared against two ablations: \begin{enumerate}
    \item a modified SMPC formulation (denoted by \eqref{opt:SMPC}{\small{$\backslash$ KF}}) that just uses \eqref{opt:gam_dyn} instead of KF for {\small{$\gamma_t$}} predictions.
    \item a modified SMPC formulation (denoted by \eqref{opt:SMPC}{\small{$\backslash\pi$}})  that just looks for an optimal sequence of control values instead of control laws, i.e., {\small{$\mathbf{u}^j_t=\mathbf{h}_t$}}.
\end{enumerate}  For each approach, both {\small{$\sigma=1,\sigma=2$}} scenarios are simulated from 16 initial conditions chosen from the set: {\small{$x_0\in\{[0\pm 1]\times[11\pm 1]\}, o_0\in\{[-9\pm 1]\times[14\pm 1]\}$}}. We deem a control task as successful if the EV is able to stop at {\small{$s_f$}}, with the TV stopping {\small{ $d_{\text{safe}}$}} behind. We record the percentage of task success as (S\%). We also record the percentage of feasibility of the SMPC optimization problems as (F\%). When infeasibility is encountered during the simulation, the previous input is applied. Table~\ref{tab:comp} presents our results for the SMPC approaches for each scenario.
\begin{table}[!ht]
% \vskip -0.1 true in
    \centering
    \caption{Ablation study for SMPC prediction approaches}
    \vskip -0.1 true in
    \label{tab:comp}    
    \begin{tabular}[c]{|c|c||c|c|c|}
        %% HEADER
        \hline
            \multirow{2}{*}{Mode}&\multirow{2}{*}{Metric} & \multicolumn{3}{c|}{SMPC}\\\cline{3-5}
        
         & &\eqref{opt:SMPC} & \eqref{opt:SMPC}{\small{$\backslash$ KF}}& \eqref{opt:SMPC}{\small{$\backslash\pi$}}\\
        \hline
        \hline
        %% ENTRIES
        \multirow{2}{*}{{\small{$\sigma=1$}}}& S\%    & $\mathbf{100.}$ & 93.75 & $\mathbf{100.}$  \\
        \cline{2-5} &
       F\%    & $\mathbf{98.71}$ & 98.21&  76.10  \\
        \hline
        \multirow{2}{*}{{\small{$\sigma=2$}}}& S\%    & $\mathbf{87.5}$ & 56.25& 0.  \\
        \cline{2-5} &
       F\%    &$\mathbf{95.30}$ & 93.38 & 8.33  \\
        \hline
    \end{tabular}
    % \vskip -0.1 true in
\end{table}
\subsection*{Discussion:}
We see that removal of feedback in the predictions results in deteriorated performance in terms of feasibility of the SMPC optimization problem and task success, especially in the scenario when interaction is required ( {\small{$\sigma=2$}} in our example).
\section{CONCLUSION}
We have proposed a Stochastic MPC framework with interaction-aware, multi-modal predictions of TVs given by basis of known features multiplied by unknown, time-varying weights. The proposed approach finds an optimal sequence of EV and TV trajectory-dependent control laws given by the affine disturbance feedback parameterization, to 1) reduce conservatism in satisfaction of chance-constraints and 2) use dual control for prediction, and estimation of feature weights.

\section*{Appendix: Matrix Definitions for Stacked Predictions}
Consider the following matrix functions:
{\vskip -0.1 true in\begingroup
\allowdisplaybreaks\small{
\begin{align*}
    &\mathcal{M}_A(\{A_k\}_{k=1}^N)=\begin{bmatrix}I& A^\top_{1}& \dots&(\prod\limits_{k=1}^{N}A_{k})^\top\end{bmatrix}^\top,\\
    &\mathcal{M}_B(\{A_k, B_k\}_{k=1}^N)=\begin{bmatrix}O&\hdots& O\\B_{1}&\hdots&O\\\vdots&\ddots&\vdots\\\prod\limits_{k=1}^{N-1}A_{k}B_{1}&\dots&B_{N}\end{bmatrix},\\
    &\mathcal{M}_D(\{A_k\}_{k=1}^N)=\text{blkdiag}(\{A_k\}_{k=1}^N).
\end{align*}
\vskip -0.1 true in
}\endgroup}
The matrices for defining the stacked predictions are obtained using these matrix functions as follows

{\begingroup
\allowdisplaybreaks
\footnotesize{\begin{align*}
    &\mathbf{A}_t=\mathcal{M}_A(\{A_k\}_{k=t}^{t+N-1}), \mathbf{B}_t=\mathcal{M}_B(\{A_k,B_k\}_{k=t}^{t+N-1}),\\
    &\mathbf{E}_t=\mathcal{M}_B(\{A_k,I\}_{k=t}^{t+N-1})\\
    &\bar{\mathbf{A}}^j_t=\mathcal{M}_A(\{\bar{A}_k+Q^j_{k|t}\}_{k=t}^{t+N-1}), \mathbf{P}^j_t=\mathcal{M}_B(\{\bar{A}_k+Q^j_{k|t},P^j_{k|t}\}_{k=t}^{t+N-1})\\
     &\mathbf{G}^j_t=\mathcal{M}_B(\{\bar{A}_k+Q^j_{k|t},G^j_{k|t}\}_{k=t}^{t+N-1}),   \bar{\mathbf{F}}^j_t=\mathcal{M}_B(\{\bar{A}_k+Q^j_{k|t},I\}_{k=t}^{t+N-1})\\
     &\tilde{\mathbf{A}}^j_t=\mathcal{M}_D(\{\bar{A}_k+Q^j_{k|t}\}_{k=t}^{t+N-1}), \tilde{\mathbf{P}}^j_t=\mathcal{M}_D(\{P^j_{k|t}\}_{k=t}^{t+N-1})\\
     &\tilde{\mathbf{G}}^j_t=\mathcal{M}_D(\{G^j_{k|t}\}_{k=t}^{t+N-1}), \mathbf{G}=\mathcal{M}_A(\{I\}_{k=t}^{t+N-1})\\
     &\mathbf{\Gamma}=\text{Unit lower triangular matrix}\\
     &\mathbf{\Gamma}^{j,\gamma}_t=\mathcal{M}_A(\{W^j_{k|t}\}_{k=t+1}^{t+N-1}),\mathbf{\Gamma}^{j,l}_t=-\mathcal{M}_B(\{W^j_{k|t},K^j_{k|t}\}_{k=t+1}^{t+N-1})\\
     &\mathbf{\Gamma}^{j,x}_t=-\mathcal{M}_B(\{W^j_{k|t},K^j_{k|t}\}_{k=t+1}^{t+N-1})\times[\mathcal{M}_D(\{P^j_{k|t}\}_{k=t}^{t+N-2})\ O]\\
     &\mathbf{\Gamma}^{j,o}_t=\mathcal{M}_B(\{W^j_{k|t},K^j_{k|t}\}_{k=t+1}^{t+N-1})\times\big(-[\mathcal{M}_D(\{\bar{A}_k+Q^j_{k|t}\}_{k=t}^{t+N-2})\ O]\\
     &~~~~~~~~~~+[\mathcal{M}_B(\{O,I\}_{k=t+1}^{t+N-1})^\top O]\big).
\end{align*}}
\endgroup
}
The matrices describing the predictions {\small{$\tilde{\mathbf{o}}_t^j$}} explicitly are given as 
{\small{\begin{align*}&\bar{\bar{\mathbf{A}}}^{j,o}_t=(\tilde{\mathbf{A}}^j_t+\tilde{\mathbf{G}}^j_t\mathbf{\Gamma}^{j,o}_t)\bar{\mathbf{A}}^j_t,\\
&\bar{\bar{\mathbf{A}}}^{j,x}_t=(\tilde{\mathbf{A}}^j_t\mathbf{P}^j_t+\tilde{\mathbf{P}}^j_t+\tilde{\mathbf{G}}^j_t(\mathbf{\Gamma}^{j,x}_t+\mathbf{\Gamma}^{j,o}_t\mathbf{P}^j_t))\mathbf{A}_t,\\ &\bar{\bar{\mathbf{B}}}^{j}_t=(\tilde{\mathbf{P}}^j_t+\tilde{\mathbf{A}}^j_t\mathbf{P}^j_t+\tilde{\mathbf{G}}^j_t(\mathbf{\Gamma}^{j,o}_t\mathbf{P}^j_t+\mathbf{\Gamma}^{j,x}_t))\mathbf{B}_t,\\
&\bar{\bar{\mathbf{G}}}^j_t=\tilde{\mathbf{G}}^j_t\mathbf{\Gamma}^{j,\gamma}_t+(\tilde{\mathbf{G}}^j_t\mathbf{\Gamma}^{j,o}_t+\tilde{\mathbf{A}}^j_t)\mathbf{G}^j_t\mathbf{G},\\
&\bar{\bar{\mathbf{F}}}^{j,n}_t=(\tilde{\mathbf{G}}^j_t\mathbf{\Gamma}^{j,o}_t+\tilde{\mathbf{A}}^j_t)\mathbf{G}^j_t\mathbf{\Gamma},\bar{\bar{\mathbf{F}}}^{j,v}_t=(\tilde{\mathbf{G}}^j_t\mathbf{\Gamma}^{j,o}_t+\tilde{\mathbf{A}}^j_t)\mathbf{F}^j_t,\\
&\bar{\bar{\mathbf{F}}}^{j,w}_t=(\tilde{\mathbf{P}}^j_t+\tilde{\mathbf{A}}^j_t\mathbf{P}^j_t+\tilde{\mathbf{G}}^j_t(\mathbf{\Gamma}^{j,x}_t+\mathbf{\Gamma}^{j,o}_t\mathbf{P}^j_t))\mathbf{E}_t,\\ &\bar{\bar{\mathbf{L}}}^j_t= (I+(\tilde{\mathbf{G}}^j_t\mathbf{\Gamma}^{j,o}_t+\tilde{\mathbf{A}}^j_t)\mathbf{F}^j_t+\tilde{\mathbf{G}}^j_t\mathbf{\Gamma}^{j,l}_t)\mathbf{l}^j_t.  
\end{align*}}}
For  {\small{$\mathbf{u}^j_t$}}, we have {\small{$\mathbf{h}_t=[h^\top_{t|t},..,h^\top_{t+N-1|t}]^\top$}} and {\small{$\mathbf{M}^w_t$}} (similarly for {\small{$ \mathbf{M}^z_t$}}) given by {\small{\begin{align*}
    \mathbf{M}^w_t=\begin{bmatrix} O & \dotsc & \dotsc & O \\
                    M^w_{t,t+1|t} & O & \dotsc & O \\
                    \vdots & \ddots & \ddots & O \\
                    M^w_{t,t+N-1|t} & \dotsc & M_{t+N-2,t+N-1|t} & O
    \end{bmatrix}.
\end{align*}}}
The matrix {\small{$\mathbf{\Gamma}^{j,z}_t$}} for defining {\small{$\mathbf{z}^j_t$}} is given by
{\small{\begin{align*}
    \mathbf{\Gamma}^{j,z}_t=\mathcal{M}_D(\{G^j_{k|t}\}_{k=t}^{t+N-1})\times\begin{bmatrix}I&\hdots&O\\W^j_{t+1|t}&\dots&O\\\vdots&\ddots&\vdots\\\prod\limits_{k=t+1}^{t+N-1}W^j_{k|t}&\dots&I\end{bmatrix}
\end{align*}}\vskip -0.1 true in}
\renewcommand{\refname}{REFERENCES}
\printbibliography
\end{document}